\documentclass[%
preprint,
 amsmath,amssymb,
 aps,
]{revtex4-2}

\usepackage{graphicx}
\usepackage{dcolumn}
\usepackage{bm}

\begin{document}

\preprint{Commentary Discussion}

\title{Can quantum Rabi model with $A^{2}$-term
  avoid no-go theorem for spontaneous SUSY breaking?
}%

\author{Masao Hirokawa}
\email{hirokawa@inf.kyushu-u.ac.jp}
\homepage{https://nvcspm.net/qstl/}
\altaffiliation{Graduate School of Information Science and Electrical Engineering, Kyushu University.}

\date{\today}

\begin{abstract}
The hierarchy problem asks why the mass of the Higgs particle 
is so much lighter than the Planck-scale mass. 
Considering the interaction of the Higgs particle
and an elementary particle in the Planck-scale,
to cope with that big difference, 
the conventional calculation 
needs the help of an arbitrary, excessive fine-tuning,
that is, the huge cancellation
between the bare mass term and the quantum correction, 
without obeying a physical principle such as symmetry.
Thus, it is often said to be unnatural.
On the other hand, the theory of supersymmetry (SUSY) is a strong candidate
naturally to solve the hierarchy problem. 
However, any sign of SUSY even for the quantum mechanics (QM) version
\cite{wit81, wit82} had not been firmly, directly observed in
the physical reality until Cai \textit{et al}. reported the observation of
the $N=2$\, SUSY and its spontaneous breaking
in a trapped ion quantum simulator \cite{cai22}
for the prototype model for SUSY QM \cite{hir15}.
In this discussion, I derive a no-go theorem for
the spontaneous SUSY breaking in the strong coupling limit
for the quantum Rabi model \cite{bra11} with the $A^{2}$-term,
and at the same time,
I show another limit proposed in the scheme
by Cai \textit{et al}. \cite{cai22}
can avoid the no-go theorem and take that model from the $N=2$\, SUSY
to its spontaneous breaking.
I propose a theoretical method to observe how the effect of $A^{2}$-term
appears in the spontaneous SUSY breaking. 
\end{abstract}

\maketitle



\newpage

In the discussion below, the annihilation and creation operators
of a $1$-mode boson are respectively denoted by $a$ and $a^{\dagger}$.
The annihilation operator $\sigma_{-}$ and the creation operator
$\sigma_{+}$ of a $2$-level atom (i.e., spin) are given by
$\sigma_{\pm}:=\frac{1}{2}(\sigma_{x}\pm i\sigma_{y})$.
The standard notation $\sigma_{x}$, $\sigma_{y}$, and
$\sigma_{z}$ are used for the Pauli matrices. 
The Hamiltonian is defined by
\begin{equation}
  H(\omega_{\mathrm{a}}, \omega_{\mathrm{c}}, g, C)
  =\frac{\hbar\omega_{\mathrm{a}}}{2}\sigma_{z}
  +\hbar\omega_{\mathrm{c}}\left( a^{\dagger}a+\frac{1}{2}\right)
  +\hbar g\sigma_{x}
  \left( a+a^{\dagger}\right)
  +\hbar Cg^{2}\left( a+a^{\dagger}\right)^{2},
  \label{eq:hamiltonian_0}
\end{equation}
where the first and second terms of RHS of equation (\ref{eq:hamiltonian_0})
are respectively the free energies of the $2$-level atom
and the $1$-mode boson.
The third term is the linear interaction between the atom and boson
with the parameter $g$ representing the coupling strength.
The last term is the quadratic interaction
$\hbar C\{g\sigma_{x}(a+a^{\dagger})\}^{2}$ with the parameter $C$
which controls the dimension and volume of the quadratic interaction energy.
This is often called $A^{2}$-term. 

It is already proved that, 
tuning the parameters, $\omega_{\mathrm{a}}$, $\omega_{\mathrm{c}}$, 
as $\omega_{\mathrm{a}}=\omega_{\mathrm{c}}=\omega$
for a positive constant $\omega$,
the Hamiltonian $H(\omega, \omega, 0, 0)$
without the $A^{2}$-term has the $N=2$\, SUSY \cite{hir15}.  
As the coupling strength $g$ gets stronger enough,
the $A^{2}$-term may appear.
Similarly to the case of the superradiant phase transition
\cite{dicke54,hl73},
a no-go theorem caused by the $A^{2}$-term \cite{rza75}
should be minded, 
and its avoidance should be argued \cite{nat10}
also for the prototype model for SUSY QM.

For every non-negative $C$,
there exists a unitary operator $U_{A^{2}}$ such that
\begin{eqnarray}
  U_{A^{2}}^{*}H(\omega_{\mathrm{a}}, \omega_{\mathrm{c}}, g, C)U_{A^{2}}
  &=&H(\omega_{\mathrm{a}}, \omega(g), \widetilde{g}, 0)   
  \label{eq:hb-trans} \\
  &=&\frac{\hbar\omega_{\mathrm{a}}}{2}\sigma_{z}
  +\hbar\omega(g)\left( a^{\dagger}a+\frac{1}{2}\right)
  +\hbar\widetilde{g}\sigma_{x}
  \left( a+a^{\dagger}\right),
  \nonumber 
\end{eqnarray}
where
$\omega(g)=\sqrt{\omega_{\mathrm{c}}^{2}+4C\omega_{\mathrm{c}}g^{2}\,}$
and $\widetilde{g}=g\sqrt{\omega_{\mathrm{c}}/\omega(g)}$.
This unitary operator $U_{A^{2}}$ is obtained in the meson pair theory
of nuclear physics \cite{hir17, hir20}.
Equation (\ref{eq:hb-trans}) is often called
the Hopfield-Bogoliubov transformation \cite{nat10}
of $H(\omega_{\mathrm{a}}, \omega_{\mathrm{c}}, g, C)$. 
The effect of the $A^{2}$-term is stuffed into
$\omega(g)$ and $\widetilde{g}$. 
We note that, for $C=0$,
the parameters satisfy $\omega(g)=\omega_{\mathrm{c}}$,
$\widetilde{g}=g$, and then,
the unitary operator is $U_{A^{2}}=I$ (i.e., the identity operator). 

For the displacement operator $D(g/\omega_{\mathrm{c}}):=
\exp\left[ g(a^{\dagger}-a)/\omega_{\mathrm{c}}\right]$,
a unitary operator is defined by
$U(g/\omega_{\mathrm{c}}):=\frac{1}{\sqrt{2}}
\left\{\left(\sigma_{-}-1\right)\sigma_{+}D(g/\omega_{\mathrm{c}})
+\left(\sigma_{+}+1\right)\sigma_{-}D(-g/\omega_{\mathrm{c}})\right\}$. 
Then, it makes the equation, 
\begin{eqnarray}
  &{}&
  U(g/\omega_{\mathrm{c}})^{*}
  \left\{
  H(\omega_{\mathrm{a}}, \omega_{\mathrm{c}}, g, 0)
  +\hbar\frac{g^{2}}{\omega_{\mathrm{c}}}
  \right\}
  U(g/\omega_{\mathrm{c}})
  \nonumber \\ 
  &=&
  \hbar\omega_{\mathrm{c}}\left(a^{\dagger}a+\frac{1}{2}\right)
    -\frac{\hbar\omega_{\mathrm{a}}}{2}
  \left\{\sigma_{+}D(g/\omega_{\mathrm{c}})^{2}
  +\sigma_{-}D(-g/\omega_{\mathrm{c}})^{2}\right\}.
  \label{eq:ut}
\end{eqnarray}

Since the arguments on the limit of Hamiltonians used below
are already established
as a mathematical method \cite{hir15, hir17, hir20}.
Thus, for simplicity,
mathematically naive arguments are made in this discussion
to explain the no-go theorem and its avoidance. 

Now we consider the strong coupling limit for
the quantum Rabi model with the $A^{2}$-term.
For instance, the strong coupling limit is
experimentally realized for the quantum Rabi model
in circuit QED \cite{yosh17} as deep-strong coupling regime \cite{cas10}. 
The parameters, $\omega_{\mathrm{a}}$, $\omega_{\mathrm{c}}$, $g$, are set
as $\omega_{\mathrm{a}}=\omega_{\mathrm{c}}=\omega$
and $g=\mathrm{g}$ for a non-negative parameter $\mathrm{g}$.
The Hamiltonian $H(\omega, \omega, \mathrm{g}, 0)$ is for
the quantum Rabi model,
and denoted by
$H_{\mbox{\tiny Rabi}}(\mathrm{g})$ for simplicity.
In the renormalization for $A^{2}$-term,
the quantities, defined by $\omega(\mathrm{g}):=
\sqrt{\omega^{2}+4C\omega\mathrm{g}^{2}\,}$ and 
$\widetilde{\mathrm{g}}:=\mathrm{g}\sqrt{\omega/\omega(\mathrm{g})}$,
are used.

\begin{figure}[htbp]
  \centering
  \vspace*{-20mm}
  \hspace*{-25mm}
  \scalebox{0.3}{
      \includegraphics{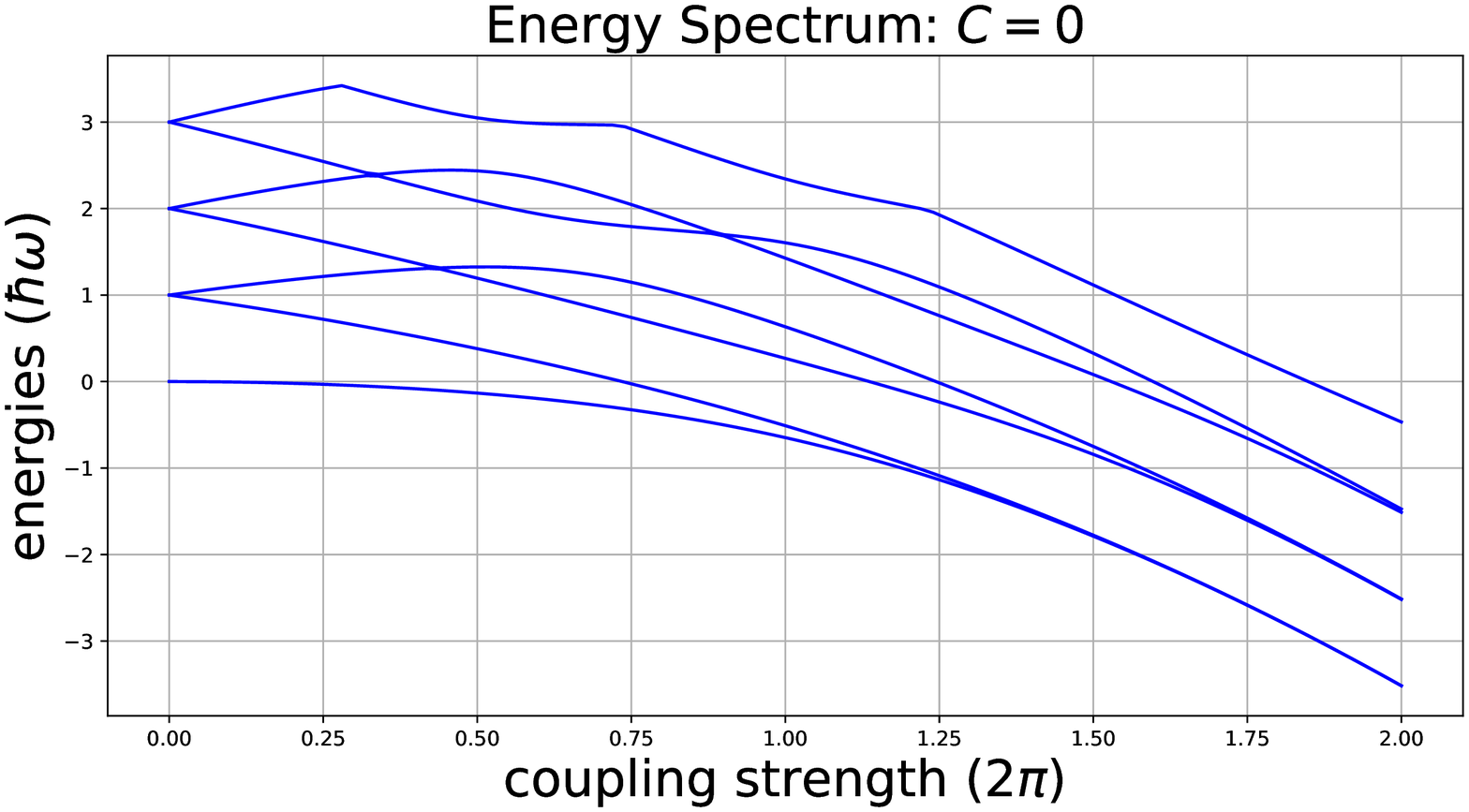}\qquad\qquad
      \includegraphics{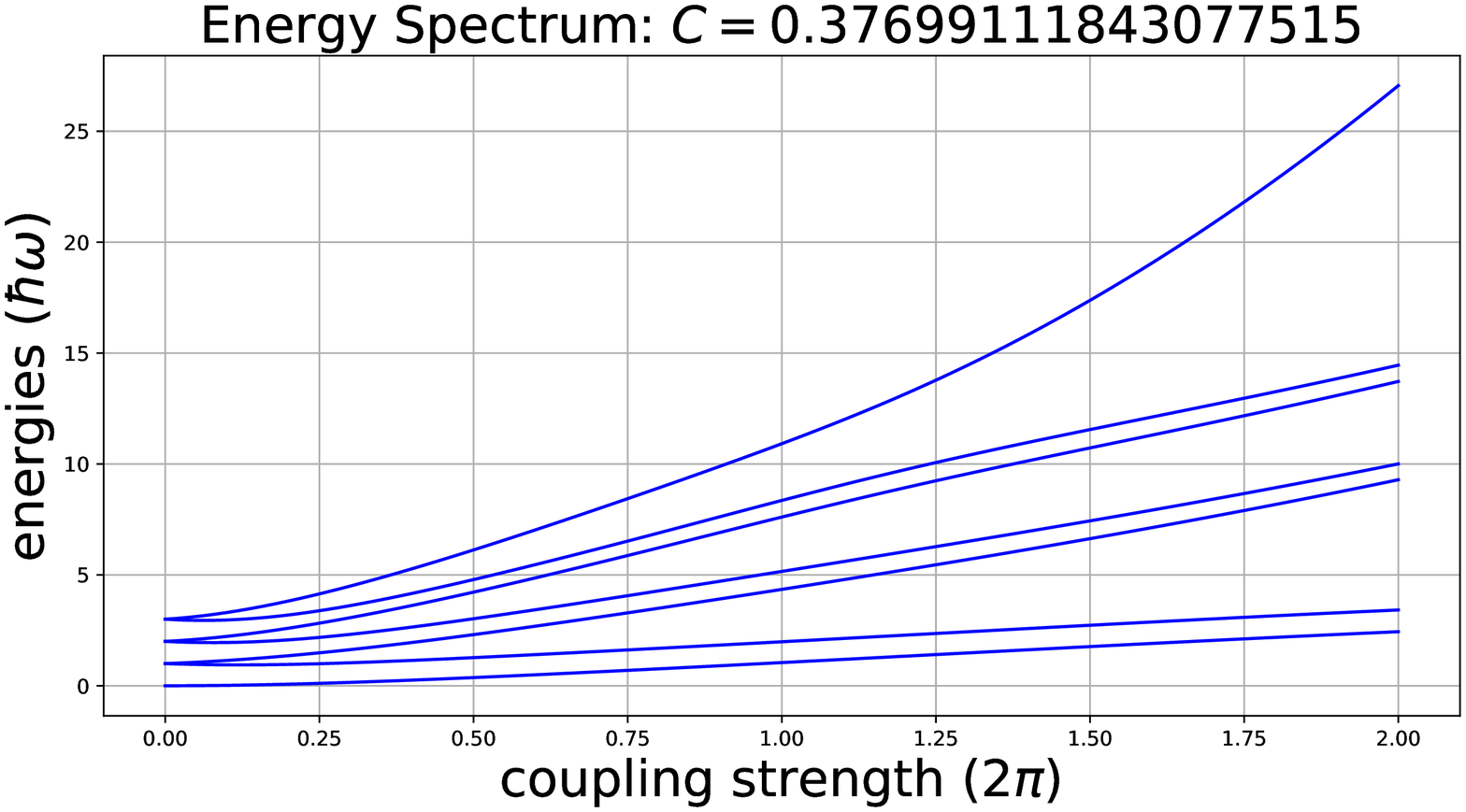}
  }
  \caption{Energy Spectrum of
    $H_{\mbox{\tiny Rabi}}(\mathrm{g})+\hbar C\mathrm{g}^{2}
    \left( a+a^{\dagger}\right)^{2}$ with
    $\omega=6.2832$:
a shows the energy spectrum for $C=0$. 
The right graph is for $C=0.3770$.
The quantum Rabi model (without $A^{2}$-term)
has the transition from the $N=2$\, SUSY to its spontaneous breaking.
On the other hand, b shows the loss of the spontaneous breaking.
Here, it should be noted $\lim_{\mathrm{g}\to\infty}\hbar\omega(\mathrm{g})=\infty$ 
and $\lim_{\mathrm{g}\to\infty}\hbar\widetilde{\mathrm{g}}^{2}/\omega(\mathrm{g})
=\hbar/(4C)$.}
  \label{fig:Fig_g}
\end{figure}

In case $C=0$, according to
the mathematical results \cite{hir15, hir17},
the approximation, 
\begin{equation}
  H_{\mbox{\tiny Rabi}}(\mathrm{g})+\hbar\frac{\mathrm{g}^{2}}{\omega}
  \approx
  U(\mathrm{g}/\omega)
  \left[
    \hbar\omega\left( a^{\dagger}a+\frac{1}{2}\right)
    \right]
  U(\mathrm{g}/\omega)^{*},
  \label{eq:approx1}
\end{equation}
is obtained as $\mathrm{g}\to\infty$.
Since the mechanism for spontaneous SUSY breaking
for the quantum Rabi model \cite{hir15}
works by the appearance of only the boson free energy in
RHS of equation (\ref{eq:approx1}),
the $N=2$\, SUSY of $H_{\mbox{\tiny Rabi}}(\mathrm{g})$
is spontaneously broken in the strong coupling limit
$\mathrm{g}\to\infty$ (Fig.\ref{fig:Fig_g}a). 
Naively, how to obtain equation (\ref{eq:approx1})
is explained in the following.
Due to the violent vibrations in the displacement operator
$D(\pm\mathrm{g}/\omega)$ as $\mathrm{g}/\omega$
grows larger,
the displacement operator decays.
Thus, the second term of RHS of equation  (\ref{eq:ut}) disappears
in the limit $\mathrm{g}\to\infty$.

In the case $C>0$, on the other hand,
the mathematical results \cite{hir20} says that 
\begin{eqnarray}
  &{}&
  H_{\mbox{\tiny Rabi}}(\mathrm{g})
  +\hbar C\mathrm{g}^{2}\left( a+a^{\dagger}\right)^{2}
  +\hbar\frac{\widetilde{\mathrm{g}}^{2}}{\omega(\mathrm{g})}
  \nonumber \\ 
  &\approx&
  U_{A^{2}}  
  U(\widetilde{\mathrm{g}}/\omega(\mathrm{g}))
  \left[
    \hbar\omega(\mathrm{g})\left( a^{\dagger}a+\frac{1}{2}\right)
    -\, \frac{\hbar\omega}{2}\sigma_{x}
    \right]
  U(\widetilde{\mathrm{g}}/\omega(\mathrm{g}))^{*}
  U_{A^{2}}^{*}
  \label{eq:approx2}
\end{eqnarray}
as $\mathrm{g}\to\infty$.
The atomic term $\hbar\omega\sigma_{x}/2$ appears in addition to
the boson free energy in RHS of
equation (\ref{eq:approx2}),
and then, this appearance interferes with the conversion to
the spontaneous SUSY breaking, and moreover, the divergnce
of $\omega(\mathrm{g})$, together with the atomic term,
rudely crushes that SUSY (Fig.\ref{fig:Fig_g}b).
Thus, the above quantum Rabi model with the $A^{2}$-term cannot go to
its spontaneous breaking as $\mathrm{g}$ chages from $\mathrm{g}=0$ to $\mathrm{g}\approx\infty$.
This is the \textit{no-go theorem} for the spontaneous SUSY breaking
in the strong coupling limit caused by the $A^{2}$-term.
The reason why $\hbar\omega\sigma_{x}/2$ appears
in RHS of equation (\ref{eq:approx2}) 
is naively explained as follows:
Since $\lim_{\mathrm{g}\to\infty}
\widetilde{\mathrm{g}}/\omega(\mathrm{g})=0$ for $C>0$
though $\lim_{\mathrm{g}\to\infty}
\widetilde{\mathrm{g}}/\omega(\mathrm{g})=
\lim_{\mathrm{g}\to\infty}\mathrm{g}/\omega=\infty$ for $C=0$,
the displacement operator
$D(\pm\widetilde{\mathrm{g}}/\omega(\mathrm{g}))$ in
equation (\ref{eq:ut}) remains as the identity operator $I$
in the limit $\mathrm{g}\to\infty$.

The approximations given by equations
(\ref{eq:approx1}) and (\ref{eq:approx2})
are mathematically established in the norm resolvent sense,
and the limit is valid over the energy spectrum \cite{rs1}. 
Thus, the limit energy spectrum is obtained by those approximations.
Whether $N=2$\, SUSY of $H(\omega, \omega, 0, 0)$ is taken
to its spontaneous breaking is checked by the energy degeneracy
between the bosonic and the fermionic states. 
The energy spectrum by the numerical analysis
with QuTiP \cite{nori1} is obtained, for instance,
as in Fig.\ref{fig:Fig_g}.
\hfill\break
(\textbf{a})\hspace*{77mm}(\textbf{b})

\vspace*{10mm}

As stated above, the Rabi model with the $A^{2}$-term is faced with
the no-go theorem in the strong coupling limit,
which is caused by the effect coming from
the $A^{2}$-term.
Indeed the no-go theorem appears in the strong coupling limit,
but there is another limit in the scheme
which is proposed by Cai \textit{et al}. \cite{cai22}.
From now on, it is proved that the limit has the advantage over
the strong coupling limit
in order that the Rabi model with the $A^{2}$-term avoid the no-go theorem
and has the spontaneous SUSY breaking.  

Let $\omega[r]$ be a continuous function of variable $r$,
$0\le r \le 1$, satisfying $\omega[0]=\omega$
and $\omega[1]=0$. 
The parameters $\omega_{\mathrm{a}}$, $\omega_{\mathrm{a}}$, $g$
are given by $\omega_{\mathrm{a}}=\omega[r]$, $\omega_{\mathrm{c}}=\omega$,
$g=r\mathrm{g}_{0}$ for a positive constant $\mathrm{g}_{0}$.
The Hamiltonian $H(\omega[r], \omega, r\mathrm{g}_{0}, 0)$
for the quantum Rabi model 
is denoted by $H_{\mbox{\tiny Rabi}}[r]$ for simplicity.
The quantities are given by 
$\widetilde{\omega}[r]:=
\sqrt{\omega^{2}+4C\omega r^{2}\mathrm{g}_{0}^{2}\,}$
and   
$\widetilde{\mathrm{g}}[r]
:=r\mathrm{g}_{0}\sqrt{\omega/\widetilde{\omega}[r]}$.
Then, the same argument as in \cite{hir15, hir17, hir20} gives   
\begin{eqnarray}
  &{}&
  H_{\mbox{\tiny Rabi}}[r]
  +\hbar Cr^{2}\mathrm{g}_{0}^{2}\left( a+a^{\dagger}\right)^{2}
  +\hbar\frac{\widetilde{\mathrm{g}}[r]^{2}}{\widetilde{\omega}[r]}
  \nonumber \\ 
  &\longrightarrow&
  U_{A^{2}}  
  U(\widetilde{\mathrm{g}}[1]/\widetilde{\omega}[1])
  \left[
    \hbar\widetilde{\omega}[1]\left( a^{\dagger}a+\frac{1}{2}\right)
    \right]
  U(\widetilde{\mathrm{g}}[1]/\widetilde{\omega}[1])^{*}
  U_{A^{2}}^{*}
  \label{eq:approx3}
\end{eqnarray}
in the norm resolvent sense \cite{rs1} as $r\to 1$. 
The naive reason why this limit is obtained is
because the limit $\omega[r]\to\omega[1]=0$
eliminates the second term of RHS of equation (\ref{eq:ut}).

\begin{figure}[htbp]
  \centering
  \vspace*{-20mm}
  \hspace*{-25mm}
  \scalebox{0.3}{
    \includegraphics{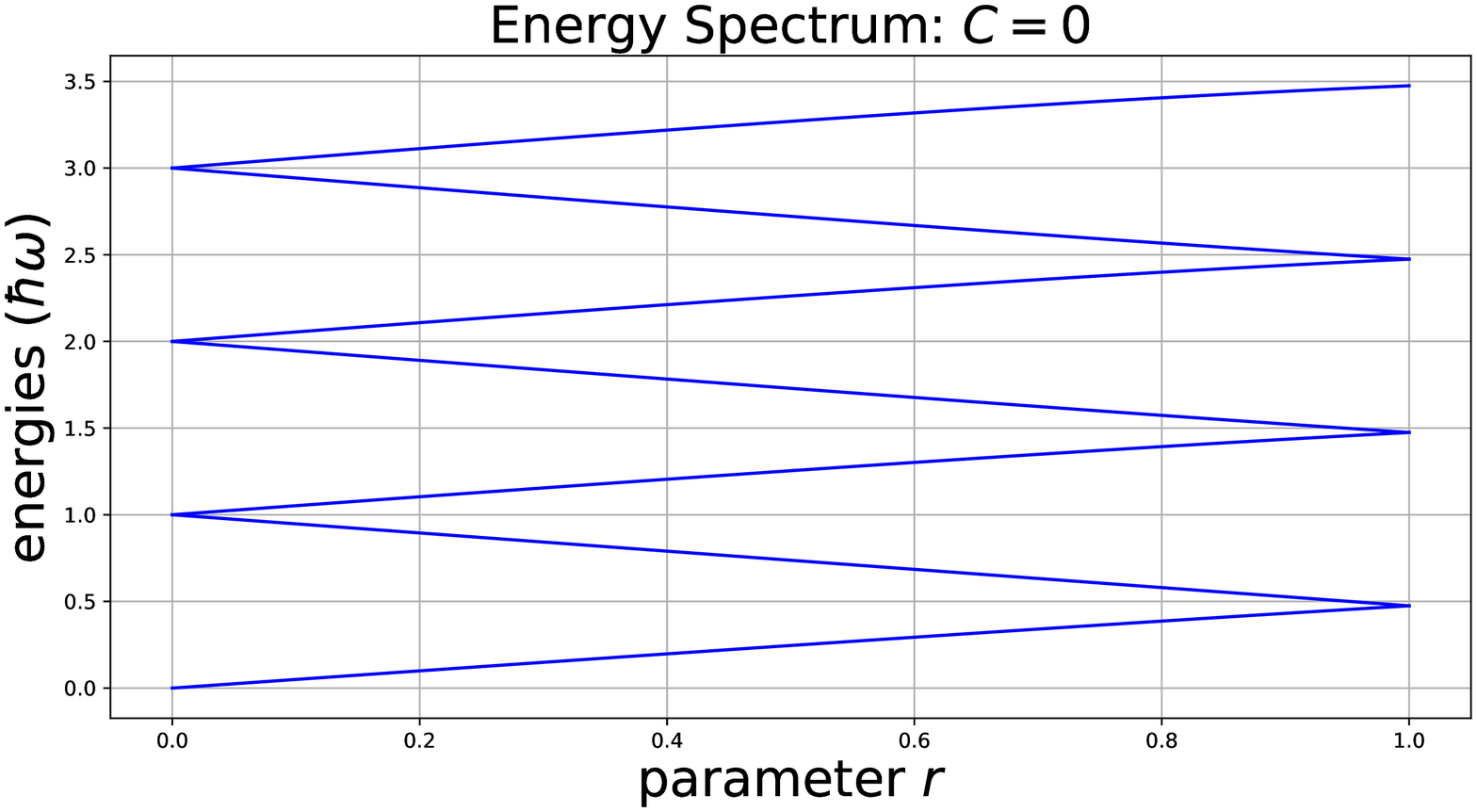}\qquad\qquad
    \includegraphics{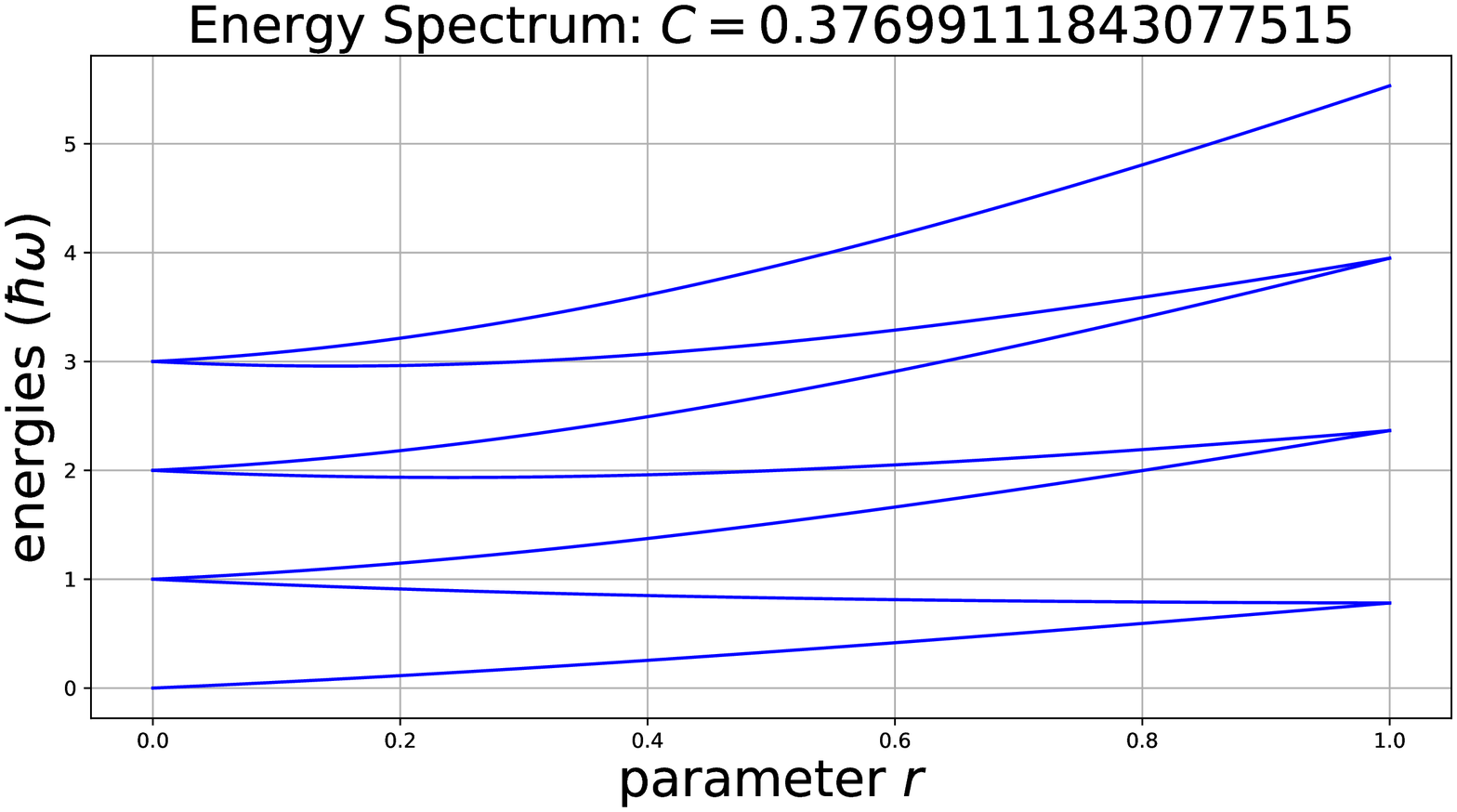}    
  }
  \caption{Energy Spectrum of
    $H_{\mbox{\tiny Rabi}}[r]+\hbar Cr^{2}\mathrm{g}_{0}^{2}
    \left( a+a^{\dagger}\right)^{2}$ with $\omega=6.2832$
    and $\mathrm{g}_{0}=6.2832$:
a shows the energy spectrum for $C=0$. 
b is for $C=0.3770$.
The quantum Rabi models without $A^{2}$-term (i.e., $C=0$)
and with $A^{2}$-term (i.e., $C>0$)  
have the transition from the $N=2$\, SUSY to its spontaneous breaking.
In particular, the energy gap by the $A^{2}$-term
appears in $\hbar\widetilde{\omega}[1]$ of the right graph.
}
  \label{fig:Fig_r}
\end{figure}

Equation (\ref{eq:approx3}) says that
the mechanism for spontaneous SUSY breaking
for the quantum Rabi model \cite{hir15} works,
and therefore, 
the Rabi model with $A^{2}$-term, described by
$H_{\mbox{\tiny Rabi}}[r]
  +\hbar Cr^{2}\mathrm{g}_{0}^{2}\left( a+a^{\dagger}\right)^{2}
  +\hbar\widetilde{\mathrm{g}}[r]^{2}/\widetilde{\omega}[r]$,
yields the spontaneous SUSY breaking in the limit $r\to 1$.
The Lagrangian for this Hamiltonian obtains the extra
second-order field, $-(m[r]^{2}/(2\hbar^{2}))\phi^{2}$, from the $A^{2}$-term, 
where $m[r]$ is a mass and $\phi=\sqrt{\hbar/(2\omega)}(a+a^{\dagger})$ 
the $1$-mode Bose field. 
We have $m[r]^{2}=4\hbar^{2}C\omega r^{2}\mathrm{g}_{0}^{2}$ then. 
The limit in the norm resolvent sense guarantees the convergence
of each energy level \cite{rs1}. 
Thus, it is worthy to note that the energy gap 
is produced in the process
from the $N=2$\, SUSY to its spontaneous breaking.
The energy gap is governed by the parameter $C$ of the $A^{2}$-term.
Whether $N=2$ SUSY of $H(\omega, \omega, 0, 0)$ goes to
its spontaneous breaking can be shown by the energy degeneracy
between the bosonic and the fermionic states. 
The energy spectrum is checked with QuTiP \cite{nori1},
for instance, as in Fig.\ref{fig:Fig_r}.
In particular, the comparison of Fig.\ref{fig:Fig_r}a and Fig.\ref{fig:Fig_r}b
shows the energy gap caused by the $A^{2}$-term.
\newpage 
\hfill\break
(\textbf{a})\hspace*{77mm}(\textbf{b})

\vspace*{20mm}

\begin{acknowledgments}
  The author acknowledges the support from
  JSPS Grant-in-Aid for Scientific Researchers (C) 20K03768.
  He would like to dedicate this study to Hiroshi Ezawa and 
  Elliott H. Lieb on the occasions of their 90th birthdays. 
\end{acknowledgments}

\bibliography{hirokawa}

\end{document}